\documentstyle[psfig,aps]{revtex}
\begin{document}
\def\v{\rule{.4mm}{2.4mm}}
\def\h{\rule[1mm]{2.4mm}{.4mm}}
\draft
\title{Self-trapped Exciton and Franck-Condon Spectra Predicted in LaMnO$_3$}
\author{Philip B. Allen and Vasili Perebeinos}
\address{Department of Physics and Astronomy, State University of New York,
Stony Brook, New York 11794-3800}
\date{\today}
\maketitle
\begin{abstract}
Because the ground state has cooperative Jahn-Teller order,
electronic excitations in LaMnO$_3$ are predicted to self-trap by
local rearrangement of the lattice.  The optical spectrum should show
a Franck-Condon series, that is, a Gaussian envelope of vibrational
sidebands.  Existing data are reinterpreted in this way.
The Raman spectrum is predicted to have strong multiphonon features.
\end{abstract}


In small molecules, excited electronic states 
generally have altered atomic coordinates, which leads
to Franck-Condon multi-phonon sidebands in electronic spectra \cite{Herzberg}.
In solids, electronic excited states are often delocalized,
eliminating such effects \cite{Sturge}.  
However, if excited states self-localize
\cite{Rashba,Song} then Franck-Condon effects should reappear in
the form of Gaussian broadening of the pure electronic
transition \cite{Cho}.  Here we argue that such effects are crucial
to a correct interpretation of LaMnO$_3$, the parent compound of
the ``colossal magnetoresistance'' materials.

\begin{figure}
\centerline{
\psfig{figure=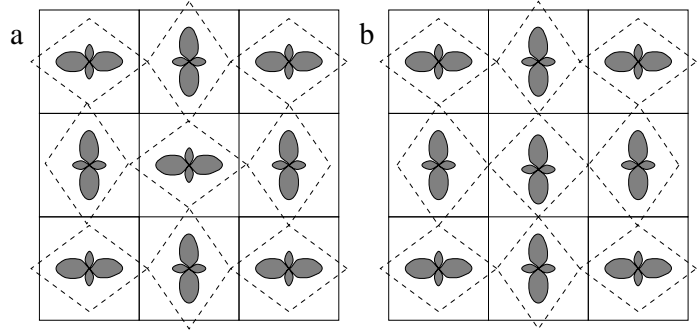,height=1.8in,width=3.685in,angle=0}}
\caption{a. schematic of Jahn-Teller distorted LaMnO$_3$ $x-y$ plane
with orbitals ordered.  b. The lowest
electronic excitation with an orbital rotated and the lattice relaxed.}
\label{states}
\end{figure}

LaMnO$_3$ has a cubic to orthorhombic cooperative JT distortion \cite{Rodriguez}
at 800K which corresponds to wavevector $\vec{Q}=(\pi,\pi,0)$.
This drives orbital order \cite{Murakami}: $x$ and $y$-oriented
$e_g$ orbitals alternate in the $x-y$ plane, in a layer structure
shown schematically in Fig. \ref{states}.
A ``minimal'' model has two Mn $d$
orbitals ($e_g$ states $\psi_{x^2-y^2}$ and $\psi_{3z^2-r^2}$),
Hubbard $U$, Hund energy $J$ to align the $e_g$ spin
with the $S=3/2$ spin of the 3 $t_{2g}$ electrons, a hopping
term permitting band formation,
and very important, electron-phonon coupling which allows
local distortions of the oxygen environment to split the
energies of the singly occupied $e_g$ levels and drive the JT transition.

The highly-doped phases of LaMnO$_3$ show a remarkable interplay of
charge, orbital, and spin order \cite{Rao}.   
The relative importance of Coulomb, electron-phonon, and other effects
is controversial.   For light doping, however, charge order is not an 
issue and the problem simplifies.  With no ``empty" (Mn$^{4+}$) sites
for an $e_g$ electron to hop into, the large size of $U$ suppresses
hopping.  The electron-phonon term becomes the dominant part
of the remaining Hamiltonian, and is simple enough that
low-temperature properties can be solved.  We have already
reported \cite{AP} properties of the self-trapped hole (``anti Jahn-Teller''
small polaron) which forms at low doping.  Its sensitivity to 
spin order influences the magnetic states seen at low doping.  We also gave a
preliminary treatment of electronic excitations \cite{Perebeinos}.
Here we give a more complete solution for the lowest electronic excitation,
finding a self-trapped exciton, insensitive to spin order, whose
Franck-Condon effect should dominate $\sigma(\omega)$ and
the resonant Raman spectrum, and might also appear in luminescence.
These results follow cleanly from our model, but have not
previously been recognized as relevant.

%


We make the strong-coupling approximation
$U\rightarrow\infty$ by considering only states with no double
occupancy of $e_g$ orbitals; this suppresses hopping 
and leaves only two terms in the Hamiltonian.  Vibrational
degrees of freedom are modelled by allowing each oxygen to move
along the nearest-neighbor Mn-O-Mn axis,
\begin{equation}
{\cal H}_{\rm vib}=\sum_{\ell,\alpha=x,y,z}(P^2_{\ell,\alpha}/2M
+Ku^2_{\ell,\alpha}/2).
\label{hel}
\end{equation}
Here $u_{\ell,x}$ is the displacement from cubic perovskite
position of the nearest oxygen in the $x$-direction
to the Mn atom at $\ell$.
The Jahn-Teller energy is modelled by a linear reduction of
energy of an occupied $3x^2-r^2$ orbital if the
corresponding two oxygens in the $\pm \hat{x}$ direction
expand outwards, and similarly for $\hat{y}$ and $\hat{z}$
oxygens if $3y^2-r^2$ or $3z^2-r^2$ orbitals are occupied,
\begin{equation}
{\cal H}_{\rm JT}=-g\sum_{\ell,\alpha} \hat{n}_{\ell,\alpha}
(u_{\ell,\alpha}-u_{\ell,-\alpha}).
\label{hJT}
\end{equation}
Here $\hat{n}_{\ell,x}$ is the occupation number of the
$3x^2-r^2$ orbital, and 
$u_{\ell,-x}$ is the displacement of the oxygen nearest
the Mn atom at $\ell$ in the $-x$-direction.
Since no hopping occurs in this model, the type and degree of
magnetic order is irrelevant, and no change occurs
in the electronic spectrum as $T$ is
increased through the Neel temperature $T_N$=140K.

The adiabatic parameter is $\alpha=\hbar\omega_0/\Delta$
where $\omega_0=\sqrt{K/M}\approx$75 meV is the oxygen vibration
frequency \cite{Raman}
and $2\Delta\approx$1.9 eV is the Jahn-Teller gap (explained below.)
Since $\alpha$ is small, we solve the Hamiltonian in
adiabatic approximation, then add quantized lattice vibrations.
There is an infinite set of equally good 
distorted ground states \cite{Millis} whose degeneracy is
broken by anharmonic terms left out of the minimal ${\cal H}=
{\cal H}_{\rm vib}+{\cal H}_{\rm JT}$.  We
simply adopt the distortion seen experimentally
and shown in Fig. \ref{states}. 
The A sublattice is defined to be Mn sites where 
$\exp(i\vec{Q}\cdot\vec{\ell})$
is 1, and the B sublattice to be sites where it is -1.  The resulting ground
state electronic wavefunction \cite{AP} is 
\begin{equation}
|0>=\prod_{\ell}^A c_X^{\dagger}(\ell)
    \prod_{\ell^{\prime}}^B c_Y^{\dagger}(\ell^{\prime})|\{0\}>,
\label{gs}
\end{equation}
where $|\{0\}>$ refers to the lattice state with oxygens displaced
by $\pm u_0 = 2g/K$
as shown in Fig. \ref{states} and in their vibrational ground states.
The orbitals $\psi_X$ and $\psi_Y$ which are occupied on the A and B
sublattices are orthogonal $x$- and $y$-directed states
$\psi_{X,Y}=(-\psi_{3z^2-r^2} \pm \psi_{x^2-y^2})/\sqrt{2}$.
In state (\ref{gs}) each occupied Mn orbital has its
energy lowered by $\Delta=8g^2/K$ and each unoccupied orbital has its
energy raised by $\Delta$.
This costs elastic energy $Ku_0^2=\Delta/2$
per Mn atom (two of the three oxygen atoms per Mn site are displaced)
giving a total condensation energy of $-\Delta/2$ per Mn.
Our picture is substantially the same as the earlier theory by Millis
\cite{Millis}.

If oxygen atoms were frozen in these optimal positions, the lowest
electronic excitation 
(an ``orbital defect,'' Fig. \ref{states} b)
would cost $2\Delta = 1.9$ eV, the JT gap.
This would form a narrow ``orbiton'' band \cite{Perebeinos} 
because terms of order $t^2/U$ (neglected here)
allow the orbital defect to exchange sites, lowering the energy by $t^2/U$.  
However, there is a much greater energy lowering if instead the lattice 
locally {\bf undistorts}, pinning the orbital excitation on a single site as
shown in Fig. \ref{states}b. This reduces the energy of the
orbital defect from $2\Delta$ to $\Delta$.  Thus the lowest electronic
excitation is a strongly self-trapped exciton, somewhat like the Frenkel 
exciton seen in molecular crystals \cite{Sturge}.  

If this excitation
is excited optically, the Franck-Condon principle applies, and 
a sequence of vibrational sidebands of the localized
exciton will appear in the spectrum.  The optical conductivity is
\begin{equation}
\sigma(\omega)=\frac{\pi N e^2 \omega}{\Omega} \sum_{n,n^{\prime}} 
\frac{e^{-\beta n_i\hbar\omega_0}}{Z}
|<fn^{\prime}|\hat{\epsilon}\cdot\vec{r}|in>|^2 
\delta(\Delta+(n_f-n_i)\hbar\omega_0-\hbar\omega),
\label{sigma}
\end{equation}\
where $\Omega$ is the volume of the crystal, and the number of
cells $N$ appears because the localized excitation can be created
on any Mn atom.  The partition function $Z$ is 
$(1-\exp(\beta\hbar\omega_0))^{-6}$ because 6 oscillators couple to each
electronic transition.  The required dipole matrix element is
\begin{equation}
<fn^{\prime}|\hat{\epsilon}\cdot\vec{r}|in>= \int d^6R \int d\vec{r}
	\chi_{n^{\prime}}(R-R_f)\psi_Y(\vec{r},R)\hat{\epsilon}_L \cdot \vec{r}
	\psi_X(\vec{r},R)\chi_n(R-R_i).
\label{metot}
\end{equation}
The crystal starts in state $|in>$
with an electron at site 0 in the electronic ground state $|i>=|X>$
with equilibrium coordinates $R_i$ and
vibrational quanta of the 6 surrounding oxygens denoted by
$\{n\}=(n_1,n_2,n_3,n_4,n_5,n_6)$.
It ends up at the same site in final state $|fn^{\prime}>$ with
the electron in state $|f>=|Y>$ of
energy $\Delta$ with new equilibrium positions $R_f$ 
and new vibrational quanta
$\{n^{\prime}\}=(n_1^{\prime},n_2^{\prime},n_3^{\prime},
                 n_4^{\prime},n_5^{\prime},n_6^{\prime})$.
The notation $n_i$ denotes the total initial vibrational level
$n_1+\ldots+n_6$, and $n_f$ is the total final vibrational level.
The electronic part of the matrix element
is a $d$ to $d$ transition which is forbidden when the surroundings
are symmetric (as in the starting coordinates $R_i$ or the final
coordinates $R_f$ if $n_i=n_f=0$).  
The origin of the observed optical transitions
has been discussed by various authors \cite{Arima,Elfimov,Ahn}.  In 
our localized picture with 
lattice displacements as primary factors, it is natural to recognize
that any asymmetric oxygen breathing displacement will cause the Mn
$e_g$ orbitals to acquire an admixture of $4p$ character. 
A typical mixing coefficient is
\begin{equation}
\gamma=\int d\vec{r} \psi_{3z^2-r^2}\frac{\partial V}{\partial u_5}
\psi_{z}/(\epsilon_d -\epsilon_p)
\label{mixing}
\end{equation}
where $\psi_z$ is an orbital of $p$-character,
partly Mn $4p$, and partly oxygen $2p$; $\partial V/\partial u_5$
is the perturbation caused by a displacement of oxygen 5 in the
$\hat{z}$-direction.  The corresponding allowed optical matrix 
element is
\begin{equation}
d=\int d\vec{r} \psi_{3z^2-r^2} z \psi_{z}.
\label{dipole}
\end{equation}
The resulting matrix element, Eqn. \ref{metot}, is
\begin{equation}
<fn^{\prime}|\hat{\epsilon}\cdot\vec{r}|in>=\frac{\gamma d}{2}
<n_1^{\prime} n_2^{\prime} n_3^{\prime} n_4^{\prime} n_5^{\prime} n_6^{\prime}|
[(u_1+u_2)\epsilon_x + (u_3+u_4)\epsilon_y -2(u_5+u_6)\epsilon_z]
| n_1 n_2 n_3 n_4 n_5 n_6 >
\label{modelme}
\end{equation}

The general term (\ref{modelme}) is complicated; we have evaluated only
when the initial state has $n_i=0$ or 1 quanta of vibration.  Two or
more quanta occur at $<$6\% probability at 300K.  Our approximate
optical conductivity is then
\begin{equation}
\sigma(\omega)=\frac{1}{Z} \left[ \sigma_0(\omega)
    + e^{-\beta\hbar\omega_0} \sigma_1(\omega) + \cdots \right]
\label{tempexp}
\end{equation}
The vibrational overlap integrals needed for $\sigma_0$ are
\begin{equation}
<n_1|0>=\left(\frac{M\omega_0}{2\hbar}\right)^{n_1/2}
\frac{(-u_0)^{n_1}}{\sqrt{n_1!}}
\exp\left(-\frac{M\omega_0 u_0^2}{4\hbar}\right)
\label{n0}
\end{equation}
\begin{equation}
<n_1|u_1|0>=\left(\frac{u_0}{2}-\frac{n_1\hbar}{M\omega_0 u_0}\right)
<n_1|0>.
\label{n1}
\end{equation}
and similar for states 2,3,4 except that for states 2 and 4, 
$-u_0$ is replaced by $+u_0$; this changes the sign of 
the prefactor of Eqn. \ref{n1}.
We find that for both $\sigma_0$ and $\sigma_1$,
$\sigma_{xx}=\sigma_{yy}=\sigma_{zz}/4$.  The answer for $\sigma_0$ is
\begin{equation}
\sigma_{0,zz}(\omega)=(\gamma d)^2\frac{\pi e^2}{M}\frac{N}{\Omega}
\frac{\omega}{\omega_0} \sum_{n=0}^{\infty} \frac{(\Delta/\hbar\omega_0)^n}{n!}
\exp\left(-\frac{\Delta}{\hbar\omega_0}\right)
\delta\left(\frac{\Delta}{\hbar}+(n+1)\omega_0 -\omega\right).
\label{sigzz}
\end{equation}
The formula for $\sigma_1$ is more complicated and will not be
given here.

\begin{figure}
\psfig{figure=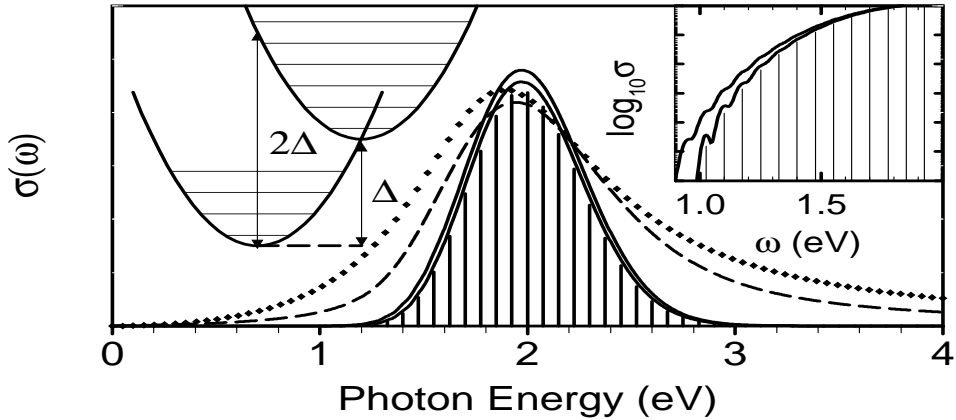,height=2.5in,width=5.4in,angle=0}
\caption{Optical conductivity of LaMnO$_3$.  The points are
the lowest Lorentzian oscillator fit by Jung {\it et al.} to
their data.  The dashed curve is a $T=0$ sum of convolved Lorentzians
centered at the vibrational replicas shown as vertical bars; the
solid curves are $T=0$ (lower) and $T=300$K (upper)
sums of convolved Gaussians, also shown in the
inset on a logarithmic scale.  Tick marks in the inset denote decades.}
\label{fc}
\end{figure}

Our theory compares reasonably well with experiment.  
$\sigma(\omega)$ has
been measured by reflectivity on polycrystalline samples at
room temperature \cite{Arima,Jung}, single crystals at low $T$
\cite{Okimoto}, and cleaved single crystals at room temperature 
\cite{Takenaka}, all with consistent results.
Jung {\it et al.} \cite{Jung} have expressed
their $\sigma(\omega)$ as a sum of Lorentzian peaks.  The lowest
peak is centered around 1.9 eV with a width of 1 eV.  Fig. \ref{fc}
shows their Lorentzian fit, and compares with our theory, with
$2\Delta=1.9$ eV and amplitude chosen to agree with experiment.

The intensity seen experimentally \cite{Jung} corresponds to
$f$=0.16 oscillators per Mn atom in the lowest peak,
where $\int d\omega \sigma(\omega)$ is defined as $(\pi N e^2/2m\Omega)f$,
with $m$ the electron mass.  Our theory gives
$f=2(\gamma d)^2(m/M)(2\Delta/\hbar\omega_0 +1)$
and has the value $0.18(\gamma d)^2\times 10^{-2}$.  To agree with experiment,
$\gamma d$ must be 10.  This is reasonable since the electron-phonon
mixing coefficient may be a few per $\AA$ oxygen displacement, while
the dipole matrix element is likely to be a few $\AA$.  Most authors
\cite{Arima,Elfimov,Ahn,Jung,Okimoto}
agree in assigning the 2 eV structure to $e_g$ to $e_g$ transitions
activated by mixing of $p$-character, usually from surrounding
oxygens.  However, the origin of the mixing is usually taken as the
dispersive broadening of the $e_g$ bands; in our strong-coupling
picture, these effects enter at order $t/U$ and are neglected,
but might be comparable to
the phonon effects which we calculate explicitly.

The delta function peaks of our model theory should
be replaced by local densities of vibrational states on oxygen atoms.
If this is mimicked by using a sequence of convolved Lorentzians
($L, L\ast L, \dots$) the result agrees closely with experiment.
A sequence of convolved Gaussians may be more realistic, and 
is shown in Fig. \ref{fc} as a dashed line and enlarged on a logarithmic scale.
Our theory is not the only candidate; band theory \cite{Solovyev} yields
reasonable agreement at $T=0$.  However, our theory makes definite
predictions which differ from those of band theory.
We find the onset of absorption at $T=0$
to be the one-phonon structure at $\Delta+\omega_{\rm ph}\approx$1 eV, 
weaker by 10$^4$ than the 2$\Delta$=1.9 eV peak
absorption.  The fine structure of subsequent phonon peaks will
be harder to resolve because of increasing multiplicity of vibrational
quanta available for multi-phonon absorption.  
Another definite prediction concerns temperature dependence.  Contrary
to approaches using dispersive bands \cite{Solovyev,Ahn}, $\sigma(\omega)$ in
our approach is not affected by the loss of magnetic order as temperature
increases.  Contrary to band theory, we predict very weak 
temperature effects near the absorption
peak at $2\Delta$, consistent with experiment \cite{Okimoto,Takenaka}.
By contrast we predict
big changes of intensity in the weak features
near $\omega=\Delta$, including new weak peaks at $\Delta-n\omega_{\rm ph}$
for $n=0,1,\ldots$ activated by temperature, and shown
in the inset to Fig. \ref{fc}.  These effects are too small to be seen
in existing reflectivity data, 
but should be measurable by absorption
in very clean thin films.
If electronic exitations survive long enough to relax by luminscence,
then a very large Stokes shift is predicted.


Another manifestation of the Franck-Condon effect should
appear in the Raman spectrum \cite{Guntherodt}, particularly if the
laser frequency is near the 2$\Delta$=1.9 eV
Jahn-Teller gap edge.  Multi-phonon Raman features
should appear with intensity similar to the one-phonon
spectrum.  The incident photon creates (among various other
virtual excitations) a self-trapped
exciton in a superposition of multi-phonon states
(as in the left inset of Fig. \ref{fc}.)  The virtual exciton
can reemit a photon, returning either to the ground state or (with
nearly equal amplitude)
to various one-phonon or multi-phonon excited states.  The theory was given
by Shorygin \cite{Shorygin}.  A particularly clear solid-state
example is Martin's \cite{Martin} observation of multiples
of the localized phonons of the MnO$_4^{2-}$ impurity complex
in CsI.  Multiples of the Jahn-Teller related vibrations
should appear in the Raman spectrum of LaMnO$_3$,
but less distinct than in Martin's work since the local
phonon density of Jahn-Teller vibrations will be less localized and
correspondingly less well-structured.  Published Raman measurements \cite{Raman}
on undoped LaMnO$_3$ do not extend to the multi-phonon region.
An unpublished two-phonon
replica of the Jahn-Teller phonon in LaSr$_2$Mn$_2$O$_7$
is seen by Romero {\it et al.} \cite{Romero}, and similar features are
seen in unpublished \cite{Romero1} work on LaMnO$_3$.  These are probably the
effect we are predicting.


\acknowledgements
We thank M. Blume, M. Cardona, J. P. Hill, T. P. Martin, 
L. Mihaly, A. J. Millis, and D. Romero for help.
This work was supported in part by NSF grant no. DMR-9725037.

\end{document}